\documentclass[tighten,twocolumn]{aastex63}

\usepackage{graphicx}
\usepackage{ulem}
\usepackage{amsmath}
\usepackage{tikz}
\usepackage{tabularx}
\usepackage{svg}

\graphicspath{{./}{figures/}}

\begin{document}

\title{Formation of Black Hole--White Dwarf X-ray Binaries in Globular Clusters}

\correspondingauthor{William Y. W. Yang}
\email{ywyang@caltech.edu}

\author[0009-0007-4064-2162]{William Y. W. Yang}
\affiliation{Division of Physics, Mathematics, and Astronomy, California Institute of Technology, Pasadena, CA 91125, USA}

\author[0000-0002-4086-3180]{Kyle Kremer}
\affiliation{Department of Astronomy \& Astrophysics, University of California, San Diego, La Jolla, CA 92093, USA}

\author[0000-0002-7444-7599]{James C.\ Lombardi, Jr.}
\affiliation{Department of Physics, Allegheny College, Meadville, Pennsylvania 16335, USA}

\author[0000-0002-8532-4025]{Kristen C.\ Dage}
\affiliation{International Centre for Radio Astronomy Research—Curtin University, GPO Box U1987, Perth, WA 6845, Australia}

\begin{abstract}
Globular clusters are host to significant populations of dynamically-active stellar remnants that connect to a variety of astrophysical sources. Using simulations performed with the \texttt{Cluster Monte Carlo} dynamics code, we study the formation of ultracompact binaries in which a stellar-mass black hole accretes material from a white dwarf companion in a sub-hour orbit. These binary systems are prime multimessenger targets, as they can be observed as both luminous X-ray sources, and as millihertz gravitational wave sources detectable by the Laser Interferometer Space Antenna (LISA). We find that black hole+giant collisions are the primary mechanism through which such systems form. We model the outcomes of these ``common envelope''-like events using the smoothed particle hydrodynamics code \texttt{StarSmasher}, and verify these collisions yield black hole+white dwarf binaries that enter Roche contact on sub-Gyr timescales via gravitational wave inspiral. We construct a mock catalog of local ultracompact X-ray sources and compare to candidate sources observed in globular clusters in the Milky Way (e.g., 47~Tuc~X9) and external galaxies (e.g., RZ~2109 in NGC~4472). Finally, we compute the gravitational wave strain for these sources, and show that of order one source may be resolvable in the Milky Way by LISA, representing a potentially powerful tool for observing new black holes in globular clusters.


\vspace{1cm}

\end{abstract}

\section{Introduction}

It is now widely established that globular clusters host robust populations of stellar-mass black holes throughout their lifetimes. For a typical cluster with roughly $10^6$ stars at birth and a \citet{Kroupa2001} initial stellar mass function, roughly $10^3$ stellar-mass black holes are expected to form via collapse of the most massive cluster stars. Once formed, these black holes quickly mass segregate to the dense center of their host cluster, undergoing dynamical interactions with one another and with cluster stars at incredibly high rates. The cumulative effect of these black hole interactions effectively ``heats'' the cluster from within via the ``binary burning'' dynamical process \citep[e.g.,][]{HeggieHut2003,BreenHeggie2013,kremer2020a}. In this case, modern numerical simulations of globular clusters demonstrate a clear correlation between the size of the cluster core and its black hole content \citep[e.g,][]{Mackey2008,Kremer2018_ngc3201,Kremer2025}. On longer timescales, the population of black holes is slowly depleted as the black holes dynamically kick one another out of their host cluster via strong gravitational interactions \citep[e.g.,][]{Spitzer1987,Kulkarni1993}. Observations of large well-resolved core radii for most old globular clusters \citep{Harris1996} indicate that most clusters still retain a moderate number (dozens to hundreds) of black holes today \citep[e.g.,][]{Askar2018,Kremer2019,Weatherford2020,Rui2021,Vitral2022,Dickson2023}.

The black hole content of globular clusters has gained significant interest in the past decade as a potential formation mechanism for gravitational wave sources \citep[e.g.,][]{PortegiesZwart2000,Rodriguez2016,Askar2017,Fishbach2017,Rodriguez2019,Kremer2019,AntoniniGieles2020,Ye2025}. In the centers of globular clusters, black holes pair up into binaries that inevitably harden as they undergo subsequent encounters with other black holes and stars in the cluster \citep[e.g.,][]{Morscher2015,Rodriguez2016}. Eventually, gravitational wave emission dominates the inspiral \citep[e.g.,][]{Peters1964}, and the black hole binary merges. Recent work has shown that a potentially significant fraction of the sources observed by the LIGO/Virgo/KAGRA detectors \citep[e.g.,][]{LIGO2016,LIGO2025} may have originated via these processes in dense clusters \citep[e.g.,][]{RodriguezLoeb2018,kremer2020b,Rodriguez2021}; however the exact contribution remains highly debated \citep[e.g.,][]{Zevin2021,MandelBroekgaarden2022}. Looking forward, black hole binaries formed in clusters have also been touted as a possible source of millihertz gravitational waves for future detectors like LISA \citep[e.g.,][]{Breivik2016,Kremer2019,LISA2023}. 

Although theoretical arguments for black holes in clusters and their potential role in forming gravitational wave sources have been around for decades \citep[e.g.,][]{Kulkarni1993,Sigurdsson1993}, observational confirmation of black holes in clusters has come only recently. Detached black hole+luminous star binaries are now confirmed in Galactic globular clusters via radial velocity observations of their stellar companions \citep{giesers2018,giesers2019}. Alongside these detached binaries, additional black hole candidates have been identified in globular clusters in relatively compact accreting configurations as low-mass X-ray binaries. These accreting black hole candidates have been observed in Galactic globular clusters in low-luminosity ($L_x \lesssim 10^{34}\,\rm{erg\,s}^{-1}$) quiescent states \citep{strader2012,chomiuk2013,millerjones2015,shishkovsky2018} as well as in extragalactic globular clusters in higher luminosity ($L_x \gtrsim 10^{38}\,\rm{erg\,s}^{-1}$) active states \citep[e.g.,][]{Kundu2002}. A subset of the latter are so-called ultraluminous X-ray sources (ULXs) with luminosities of $10^{39}\,\rm{erg\,s}^{-1}$ or more \citep[e.g.,][]{Maccarone2007}. With inferred accretion rates comparable to or in excess of the classic Eddington limit for a stellar-mass compact object, these ULXs have also been touted as possible evidence for intermediate-mass black holes with masses in excess of $10^2\,M_{\odot}$ which have long been speculated to reside in globular clusters \citep[e.g.,][]{Irwin2010,greene2020,Wiktorowicz2025}. In recent years, a growing number of ULXs have been identified in globular clusters of nearby galaxies \citep[e.g.,][]{Dage2019,dage2020,dage2021}, demonstrating the utility of these sources as a means of constraining the presence and dynamical evolution of black holes in globular clusters.

Detailed theoretical investigations into the formation pathways and properties of these accreting black hole binaries are crucial for interpreting these various observed sources and for investigating the broader implications for forming gravitational wave sources. Previous studies have demonstrated that low-mass X-ray binaries can form in globular clusters through a variety of dynamical processes including tidal capture \citep[e.g.,][]{Fabian1975}, exchange interactions \citep[e.g.,][]{SigurdssonPhinney1995}, physical collisions \citep[e.g.,][]{Verbunt1987,Lombardi2006}, and secular dynamics induced by the presence of a tertiary companion \citep[e.g.,][]{ivanova2010}. In principle, donor stars on the giant branch, main sequence, or even white dwarfs are all viable and would all produce unique orbital and spectral signatures. Although previous $N$-body simulations \citep[e.g.,][]{kremer2020b} have demonstrated that low-mass X-ray binaries with main-sequence/giant donors are most commonly formed, relatively compact systems with white dwarf donors may survive much longer in the dense core of a cluster under the influence of incessant dynamical encounters that constantly break apart black hole binaries \citep[e.g.,][]{HeggieHut2003}. A handful of extragalactic ULXs have been linked to accretion from white dwarf donor stars via observations of [OIII] spectral lines \citep[e.g.,][]{Zepf2007}. The Galactic black hole candidate 47~Tuc~X9 in the globular cluster 47 Tucanae exhibits spectral features that also hint at a carbon-oxygen white dwarf donor \citep[e.g.,][]{millerjones2015}. Furthermore, 28-minute modulations in the X-ray lightcurve of this source hint at an ultracompact orbit \citep{bahramian2017}. At millihertz orbital frequencies, black hole+white dwarf binaries like 47~Tuc~X9 may potentially be observable as gravitational wave sources, making them a unique multimessenger source for detectors like LISA \citep{LISA2023}.

In this paper, we study the formation of these binaries in which a stellar black hole accretes material from a white dwarf companion, henceforth referred to as ultracompact X-ray binaries (UCXBs). In Section~\ref{sec:rate}, we explore the cluster models from the \texttt{CMC Cluster Catalog} \citep{kremer2020b} to determine the main formation channels for UCXBs. In Section~\ref{sec:cluster_properties}, we describe how the rate of black hole+giant collisions (identified as our primary formation mechanism) varies with cluster properties. In Section~\ref{sec:hydro}, we perform hydrodynamic simulations of black hole+giant collisions using \texttt{StarSmasher} to study the post-collision orbital properties of the binaries and verify the viability for producing an ultracompact binary that begins mass transfer within a Hubble time. In Section~\ref{sec:MT_Xray}, we model the subsequent mass transfer for these UCXBs and estimate their X-ray luminosities. In Section~\ref{sec:mocksample}, we build a mock catalog of Galactic and extragalactic globular clusters to estimate the number of observable systems at present. In Section~\ref{sec:GWs}, we discuss the gravitational wave signal of these sources, and estimate potential detectability for LISA. We discuss our results and conclude in Section~\ref{sec:conclusions}.

\section{Formation channels in globular clusters}
\label{sec:rate}

Following previous studies on this topic \citep[e.g.,][]{ivanova2010} we consider four primary formation channels through which black hole+white dwarf binaries can form dynamically in dense stellar environments: \textbf{(1) exchange interactions} where the binary is assembled during a multi-body encounter; \textbf{(2) giant collisions} where a black hole physically collides with a giant branch star leading to a common envelope-like event in which the black hole and giant core spiral within the giant envelope, ultimately leaving behind a compact binary that evolves to become a black hole+white dwarf; \textbf{(3) tidal capture} in which a black hole passes sufficiently close to a white dwarf (within a few radii) to inject orbital energy into the white dwarf via tidal interaction, leading to formation of a (highly-eccentric) black hole+white dwarf binary; \textbf{(4) triple-induced mass transfer} where the eccentric Kozai-Lidov mechanism \citep[e.g.,][]{Naoz2016} induces mass transfer from the white dwarf to the black hole in the inner binary of a hierarchical stellar triple. As described in \citet{ivanova2010}, a combination of these scenarios may also play a role. For example, secular interactions of a stellar triple could lead to merger of a black hole+giant inner binary, ultimately leading to formation of a black hole+white dwarf binary via a common envelope event similar to scenario (2).


We examine a set of \textit{N}-body simulations of globular clusters with a range of parameters to explore the rates of these formation channels for accreting black hole+white dwarf systems. We first describe our models, then discuss each of these four channels.

\subsection{Globular cluster models}
\label{sec:methods}

To examine the dynamical processes through which interacting black hole+white dwarf binaries form within globular clusters, we use the 148 \textit{N}-body simulations from the \texttt{CMC Cluster Catalog} \citep{kremer2020b} performed using the \texttt{Cluster Monte Carlo} (\texttt{CMC}) code, a H\'enon-type Monte Carlo code for stellar dynamics \citep{rodriguez2022}. \texttt{CMC} simulations model large-scale cluster behavior, binary stellar evolution using the binary synthesis code \texttt{COSMIC} \citep{breivik2020}, and directly integrate small-\textit{N} encounters \citep{Fregeau2004}, enabling us to study in detail the various potential formation pathways for UCXBs.

As summarized in detail in \citet{kremer2020b}, the initial virial radius ($r_v=[0.5,1,2,4]\,$pc), metallicity ($Z=[0.01,0.1,1]\times Z_\odot$), initial number of cluster objects ($N=[2,4,8,16,32]\times10^5$), and galactocentric distance ($R_{\rm gc} = [2,8, 20]\,$kpc) of our simulations have been varied such that the models collectively cover the observed parameter space for the Milky Way globular clusters. We adopt a \citet{Kroupa2001} initial stellar mass function ranging from $0.08-150\,M_\odot$. All models adopt an initial binary fraction of $5\%$ across all masses, with initial orbital periods, eccentricities, and mass ratios drawn as in \citet{kremer2020b}. 

Finally, since we wish to produce predictions and comparisons to present-day observations, we will consider only late-time events within the simulation occurring later than $9\,$Gyr, representative of ages for old globular clusters. Henceforth, we use the term ``late times" to refer to $t>9\,$Gyr in our models, unless otherwise indicated.

\vspace{1cm} 

\subsection{Exchange interactions}

\begin{figure}
    \centering
    \includegraphics[width=1.0\linewidth]{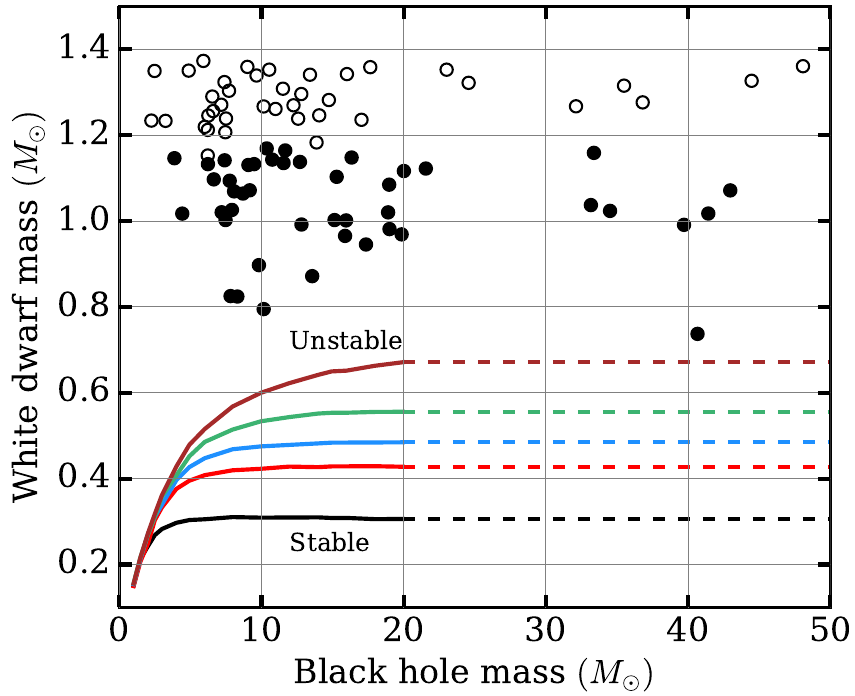}
    \caption{White dwarf mass versus black hole mass (at onset of mass transfer) for all interacting binaries formed at late times via exchange encounters in our \texttt{CMC} models. Open and filled circles denote white dwarfs of O/Ne/Mg and C/O composition, respectively. The five colored curves mark the boundaries for stable mass transfer from \citet{Church2017}, for several different assumptions concerning the mean common envelope radius and super-Eddington accretion. The dashed segments denote extrapolation of the \citet{Church2017} analysis, which only explore black hole masses up to $20\,M_{\odot}$. Even under the most optimistic assumptions, we predict all interacting binaries formed via exchange encounters are dynamically unstable and lead to a merger.\label{fig:bhwd_exchange}}
\end{figure}


Exchange encounters are a canonical pathway for dynamically assembling new binaries in dense stellar environments. In this scenario, a binary encounters either a single star (or another binary), and one or more of the objects in the encounter exchange partners \citep[for review, see][]{Fregeau2004}. Previous studies have shown exchange encounters may play an important role in the formation of X-ray binaries in stellar clusters \citep[e.g.,][]{SigurdssonPhinney1995,ivanova2010,Kremer2018_xrb}. The \texttt{CMC Cluster Catalog} tracks all strong gravitational encounters and collisions between \mbox{few-body} systems. We use these records to identify all instances where a new black hole+white dwarf binary is assembled via an exchange interaction and goes on to enter Roche lobe overflow. In practice, Roche lobe overflow is enabled in these binaries when gravitational wave emission drives the binary to smaller orbital separations.

In our full set of \texttt{CMC} models, we identify 84 black hole+white dwarf binaries formed via exchange that enter Roche contact at late times ($t>9\,$Gyr). We plot the white dwarf mass versus black hole mass for all of these systems in Figure~\ref{fig:bhwd_exchange}. As shown, the systems assembled via exchange are heavily biased toward the most massive white dwarfs; all but one of our systems have a white dwarf mass of at least $0.8\,M_{\odot}$. This is a result of mass segregation. Only the most massive white dwarfs are able to sink and dynamically interact with the relatively-massive black holes at appreciable rates. The average cluster white dwarfs (average mass of roughly $0.6\,M_{\odot}$) are, in general, relegated to the cluster outskirts and generally interact with other low-mass stars \citep[see also][]{Kremer2021_wd}. Of our exchange sample shown in Figure~\ref{fig:bhwd_exchange}, 64\% feature a white dwarf of oxygen-neon-magnesium (O/Ne/Mg) composition, and the remaining 36\% are massive carbon-oxygen (C/O) white dwarfs.

Previous studies have examined the dynamical stability of ultracompact binaries where a white dwarf fills its Roche lobe and accretes onto a compact object \citep[e.g.,][]{Nelemans2001,Yungelson2002,vanHaaften2012,Metzger2012,Bobrick2017,Church2017,tauris2018}. Of key relevance to our work, \citet{Church2017} explore this topic in the context of the 47~Tuc~X9 ultracompact binary candidate \citep{bahramian2017}. This study shows that, even allowing for highly optimistic binary evolution assumptions concerning super-Eddington accretion and the common envelope radius, mass transfer in ultracompact binaries with white dwarf masses larger than roughly $0.7\,M_{\odot}$ is always dynamically unstable and leads to a merger. We show as colored curves in Figure~\ref{fig:bhwd_exchange} the stability boundaries derived in \citet{Church2017} for various assumptions. Solid curve segments show the boundaries directly derived in their study, and dashed segments denote simple extrapolations to higher black hole masses (only 18\% of our exchange binaries have black hole masses larger than $20\,M_{\odot}$, so the details/validity of this extrapolation do not significantly impact our conclusions here).

Based on the results from \citet{Church2017}, we conclude that \textit{all} ultracompact binaries formed via exchange in our models ultimately lead to merger, not stable mass transfer. These mergers may lead to interesting astrophysical transients in their own right \citep[for example, see][]{Metzger2012,Bobrick2022} but likely do not lead to a long-lived X-ray source. This result is distinct from the conclusions of previous studies that showed exchange encounters may be important \citep[e.g.,][]{ivanova2010}. These previous studies used semi-analytic techniques to model the dynamical encounters expected within a typical cluster core. This demonstrates that full $N$-body-style simulations that self-consistently include effects like mass segregation are necessary to understand this process more fully.


\begin{figure}
    \centering
    \includegraphics[width=1\linewidth]{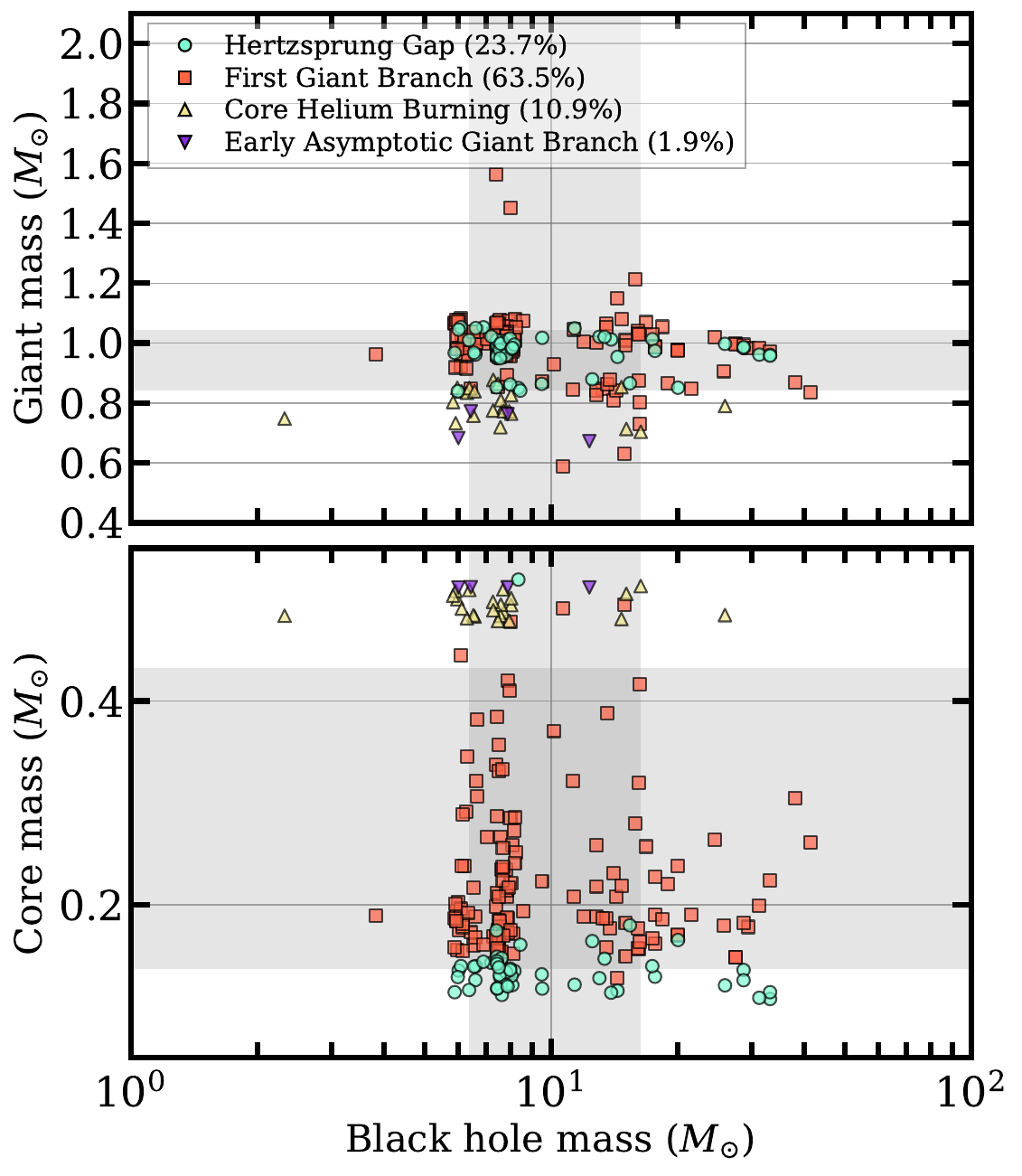}
    \caption{Giant star and their respective core masses versus black hole mass for all late-time black hole+giant collisions in the \texttt{CMC Cluster Catalog} models. The gray bands indicate the $70^\text{th}$ percentiles of each distribution. We find a median median giant mass of $0.98\,M_{\odot}$, a median core mass (bottom panel) of $0.19\,M_{\odot}$, and a median black hole mass of $7.82\,M_{\odot}$. The various colors and symbols denote the evolutionary stage of the giant with their respective abundances shown in the legend.\label{fig:BHgiantmass}}
\end{figure}

\subsection{Black hole+giant collisions}

Second, we explore formation of ultracompact binaries via collisions of black holes and giant stars. In \texttt{CMC}, we record all instances where a black hole collides with a star (main sequence stars, giants, or white dwarfs). In total, we identify $6,956$ black hole+giant collisions in our full set of models across all times. We can anticipate that most collisions relevant for producing a present-day UCXB will have occurred at late times (we address these timescales directly in Section~\ref{sec:MT_Xray}). Of our full population of $6,956$ collisions, $404$ occur at late times ($t>9\,$Gyr). For our sample of 126 \texttt{CMC} models that survive to present-day \citep[see][]{kremer2020b}, this corresponds to a collision rate per cluster of $\mathcal{R} \approx 6\times10^{-10}\,\rm{yr}^{-1}$ at late times. Assuming all of these black hole+giant collisions successfully lead to formation of a UCXB (we examine this assumption more rigorously in Section~\ref{sec:hydro}), we can take $\mathcal{R}$ to correspond roughly to the formation rate of UCXBs.

We show all black hole+giant collisions occurring at late times in Figure~\ref{fig:BHgiantmass}. The top panel shows the giant mass versus black hole mass at time of collision and the bottom panel shows the core mass of the giant versus black hole mass. The gray bands denote the $70^{\rm th}$ percentile values for black hole, giant, and core masses around the median. As shown, our median event is a roughly $8\,M_{\odot}$ black hole colliding with a $1\,M_{\odot}$ giant that has a $0.2\,M_{\odot}$ core.

As expected given the restriction to $t>9\,$Gyr, these giants are all low-mass stars with masses of roughly $1.1\,M_{\odot}$ or less. There is also a smaller population of more massive giants ($\gtrsim 1.5\,M_{\odot}$) that are stellar-collision products. We neglect this population of giants because their core masses cannot be obtained via simple stellar evolution from an initial state, and we estimate they would produce less than order unity corrections to our final numerical estimates. With different colors and marker shapes, we denote the evolutionary stage of the giants: horizontal branch, first giant branch, core helium burning, or asymptotic giant branch. Given that the lifetime of stars on the horizontal branch and first giant branch is much longer than the other two stages, collisions involving giants that have yet to begin helium core burning are much more common. This is consistent with results from previous studies \citep[e.g.,][]{ivanova2005,Lombardi2006,ivanova2010}.

We expect that if the envelope is stripped via collision from a giant yet to begin helium-core burning, the end result is a low-mass helium white dwarf. For collisions of giants that have already begun helium core burning, we expect the core to continue to ``burn out'' even after the envelope is stripped during the collision, ultimately leading to formation of a C/O white dwarf. Critically, the core mass of all giants identified here lies below the boundary for stable mass transfer identified in \citet{Church2017} and shown in Figure~\ref{fig:bhwd_exchange}. In this case, assuming the black hole+white dwarf binary formed following the collision is compact enough to ultimately come into Roche contact via gravitational wave emission (we explore this point in detail in Section~\ref{sec:hydro}), then the resulting mass transfer will be stable and potentially produce a viable long-lived X-ray source.

\begin{figure*}
    \centering
    \includegraphics[width=0.8\linewidth]{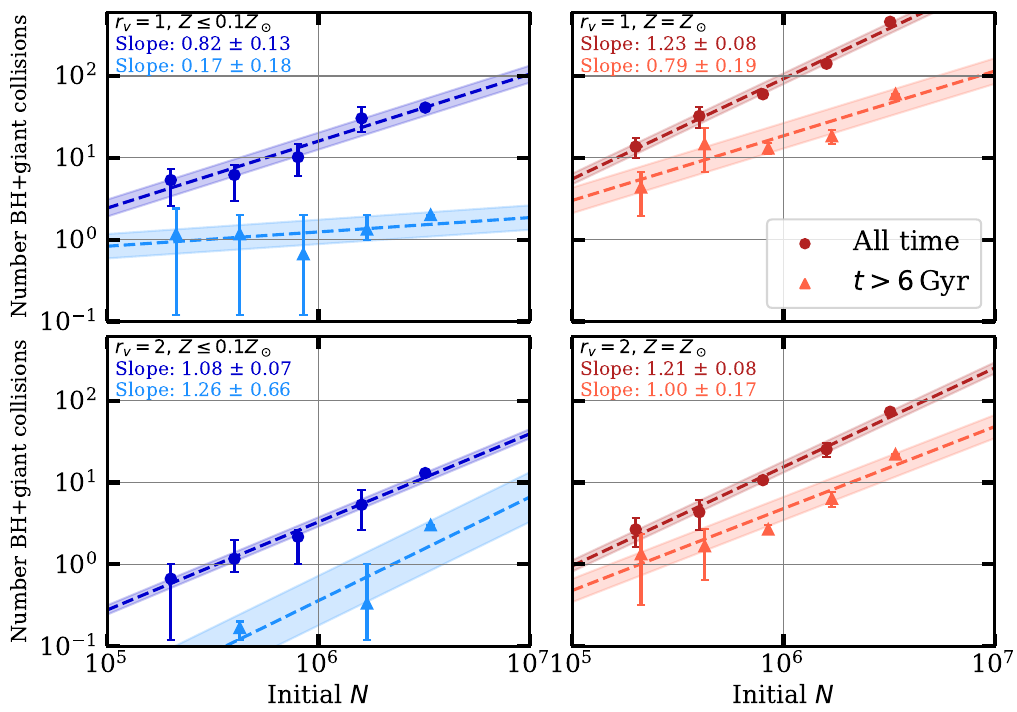}
    \caption{Number of black hole giant collisions versus initial cluster $N$ for all \texttt{CMC Cluster Catalog} models. The top (bottom) two panels show results for models with $r_v=1\,$pc ($r_v=2\,$pc), and the left (right) panels show results for low (high) metallicity models. Dark scatter points show results for all collisions throughout complete lifetime of the models, and light points restrict to only the collisions occurring at late times, here defined as $t>6\,$Gyr. The dashed curves show our best-fit linear correlation for each group of points. We also indicate the best-fit slope for each group in the captions of each sub-panel.\label{fig:correlation_N}}
\end{figure*}

\subsection{Tidal capture}

Although not examined in previous studies, we consider here a another potential scenario for forming an ultracompact black hole+white dwarf binary: via tidal capture. In this case, a black hole and white dwarf undergo a sufficiently close passage (either during a single-single encounter or during a binary-mediated resonant encounter) to form a bound pair by injecting orbital energy into the white dwarf via tidal interaction \citep[e.g.,][]{Fabian1975,PressTeukolsky1977}. Given the very close pericenter distances required (of order the white dwarf radius), we can speculate this will form a compact and highly eccentric binary that will quickly circularize and enter Roche contact via gravitational wave driven inspiral \citep[for discussion of a related scenario in the context of more massive black holes, see][]{Ye2023}. Although a careful treatment of the critical cross section for tidal capture is beyond the scope of this study, we can reasonably estimate that it will be a small multiple of the white dwarf radius \citep[in the gravitational focusing regime, the cross section scales linearly with radius; e.g.,][]{LeeOstriker1986}. In this case, the total number of physical collisions between white dwarfs and black holes (which is recorded in our \texttt{CMC} models) is a good proxy for the tidal capture rate.

In our models, we identify 24 such collisions in total at late times. Assuming that the cross section for tidal capture is roughly three times the cross section for physical collision, we can estimate of order $100$ tidal captures, roughly comparable to our predicted rate of interacting black hole+white dwarf binaries formed via exchange (Figure~\ref{fig:bhwd_exchange}). However, as in the exchange scenario, \textit{all} of the black hole+white dwarf collisions identified in our \texttt{CMC} models have white dwarf mass of roughly $0.7\,M_{\odot}$ or more. Therefore, even if tidal capture can successfully form an interacting binary of this type, we expect the mass transfer to be dynamically unstable and fail to form a long-lived accreting system. In this case, black hole+giant collisions remain as the only viable pathway for forming a long-lived UCXB.

\subsection{Triple-induced mass transfer}

\citet{ivanova2010} shows that mass transfer induced via the eccentric Kozai-Lidov mechanism in hierarchical triples \citep[e.g.,][]{Naoz2016} may play an important (and perhaps even dominant) role in the formation of UCXBs. The triple mechanism could in principle enhance the formation rate of all three of the scenarios discussed above. However in practice, the exchange and tidal capture scenarios would still be limited by the mass transfer stability arguments of \citet{Church2017}. In those cases, although secular interactions in triples may enhance the rate, they still would be unlikely to form long-lived accreting X-ray sources. However, triples may indeed enhance the formation rate via black hole+giant collisions, since the mass transfer instability concern does not then apply.

Although we do not model secular interactions in triples within \texttt{CMC}, we can explore the possible effect in post-processing by examining all hierarchical triples formed in our simulations during binary-mediated encounters. \citet{Fragione2020_CMCtriples} performed this exercise for all triples formed in the \texttt{CMC Cluster Catalog} models considered here. That study found (see their Table A1) that $238$ total triples formed across all \texttt{CMC} models with a black hole+giant inner binary (compared to the population of roughly $7,000$ black hole+giant collisions). Assuming optimistically that secular interactions drive \textit{all} of these inner binaries to merge, this still is a minor correction to the total black hole+giant collision rate. In this case, we conclude that although secular effects in triples may increase the black hole+giant collision rate slightly, it is unlikely to dominate the rate. As described in later sections, we find that additional effects from triples are not required to explain the observed rates of UCXBs in clusters: standard black hole+giant collisions are sufficient.

Motivated by our key result that black hole+giant collisions appear to be the dominant pathway through which UCXBs may form, we now go on to explore these specific interactions in further detail.

\vspace{1cm}

\section{Dependence of collision rate on cluster properties} 
\label{sec:cluster_properties}

In Figure~\ref{fig:correlation_N}, we show the relationship between the number of black hole+giant collisions with cluster mass (here encoded as the initial $N$ value for the simulations; see Section~\ref{sec:methods}). The top two panels show correlations for models with $r_v=1\,$pc, and the bottom two for $r_v=2\,$pc. The left panels (blue) show correlations for our low-metallicity models ($Z=[0.01,0.1]\,Z_{\odot}$) and right (red) for high-metallicity ($Z=Z_{\odot}$). In each panel we distinguish between the total number of black hole+giant collisions occurring throughout the entire history of the model (shown as the darker color), and only those collisions that occur at late times (defined here as $t>6\,$Gyr to increase the sample size of events), that are more representative of the types of collisions that may form UCXBs in present-day clusters. Each scatter point includes the average of all available \texttt{CMC Cluster Catalog} models that meet the criteria of each panel. For example, the dark blue scatter point at $N=2\times10^5$ in the top-left panel averages the result from six different \texttt{CMC} models (two metallicities and three different values for $R_{\rm gc}=[2,8,20]\,$kpc). We assume galactocentric distance plays a negligible role on the number of black hole+giant collisions, which in general all occur in their host cluster's core that is not heavily influenced by the cluster's outer tidal boundary. The error bars for each scatter point denote the $1\sigma$ range for the group of models considered.

\begin{figure*}
    \centering
    \includegraphics[width=0.8\linewidth]{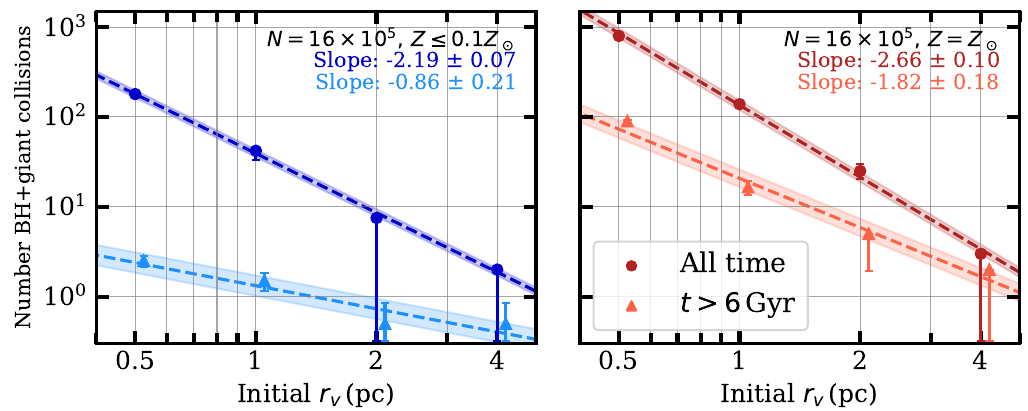}
    \caption{Same as Figure~\ref{fig:correlation_N}, but now showing the number of black hole+giant collisions versus initial cluster virial radius, $r_v$ for all \texttt{CMC} models.\label{fig:correlation_rv}}
\end{figure*}

The dashed curves denote the best-fit linear correlation identified for each group of points, with the error bars denoting the $1\sigma$ error for each correlation. Not surprisingly, the uncertainty is larger when we restrict to only late-time mergers (there are fewer, and subject to small number statistics, particularly for lower metallicities). We also print the best-fit slope value for the curves in each subpanel.

As shown in the figure, higher-metallicity clusters clearly produce more black hole+giant collisions, both overall and at late times. This enhancement is roughly a factor of ten, across all cluster masses. Higher-metallicity clusters generally form lower-mass black holes \citep[e.g.,][]{kremer2020b,Ye2025}, as a result of enhanced mass loss via line-driven stellar winds \citep[our \texttt{CMC} models use the stellar-wind prescriptions of][]{Vink2001}. Higher-mass black hole populations formed at low metallicities are more dynamically decoupled from the relatively low-mass stellar populations in their host \citep[e.g.,][]{kremer2020b}, therefore reducing the dynamical interaction rate between the black hole and stellar (giant) populations. Meanwhile, lower-mass black hole populations dynamically mix more with stars, leading to an enhancement in the collision rate. We discuss this apparent trend with metallicity further in Section~\ref{sec:observed_UCXB}.

We find that the number of black hole+giant collisions scales roughly linearly with cluster mass, $N_{\rm BHG} \propto M_{\rm cl}$, although the scaling appears to be slightly steeper at higher-metallicities. For comparison, \citet{Mai2026} shows that the total number of binary black hole mergers scales with cluster mass as roughly $N_{\rm BBH} \propto M_{\rm cl}^{1.4}$, slightly steeper than the slopes identified here.

In Figure~\ref{fig:correlation_rv}, we show number of black hole+giant collisions versus cluster radius, parameterized here by the initial $r_v$. Again the left and right panels (blue and red) show results for our low and high metallicity models. As in Figure~\ref{fig:correlation_N}, each scatter point indicates the average of all \texttt{CMC} models for various $r_v$. We show results here only for our $N=1.6\times10^6$ \texttt{CMC} models, as these are the only set with enough black hole+giant collisions across the full range of $r_v$ to yield meaningful results. As expected, denser clusters (lower $r_v$) feature an enhanced collision rate. As a rough approximation, we find $N_{\rm BHG} \propto M_{\rm cl} r_v^{-2}$ provides a good proxy for the scaling exhibited by our models. We use this scaling in Section~\ref{sec:mocksample} to construct a mock local Universe UCXB sample.

\begin{deluxetable*}{l | c c c c c | c c c c | c}
\tablecaption{Initial masses and initial pericenter distances for the four SPH simulations performed, as well as final giant masses, semi-major axes, eccentricities, and gravitational wave inspiral times.\label{table:SPH}}
\tablehead{
\colhead{} & 
\multicolumn{5}{|c|}{Initial properties} &
\multicolumn{4}{c|}{Final properties} &
\colhead{video}
}
\startdata
 & $M_{\rm bh}$  & $M_{\rm G}$ & $M_c$ & $R_{\rm G}$ & $r_{\rm p}/R_{\rm G}$ & $M_{{\rm G},f}$  & $a_f$ & $e_f$ & Inspiral time & \\ 
& \(\,(M_\odot)\) & \(\,(M_\odot)\) & \(\,(M_\odot)\) & \(\,(R_\odot)\) &  & \(\,(M_\odot)\) & \(\,(R_\odot)\) & & (Myr) & \\
\hline
\texttt{HGrpRg0.5} & 10 & 0.85   & 0.137 & 2.2  & 0.5  & 0.155   & 4.59   & 0.842 & 60 & \href{https://drive.google.com/file/d/1OSL6BqAqGWNaRxTvndZZzou3DlkUPI3o/view?usp=drive_link}{link} \\ 
\texttt{HGrpRg0.75} & 10 & 0.85   & 0.137 & 2.2  & 0.75  & 0.173   & 7.41   & 0.851 & 300 & \href{https://drive.google.com/file/d/1Sk4O-Ir4C7HgJdvuejzuk1NesjQkCdig/view?usp=drive_link}{link}  \\ 
\texttt{FGBrpRg0.5} & 10 & 0.85   & 0.186 & 4.0  & 0.5  & 0.205   & 12.3  & 0.885 & 870 & \href{https://drive.google.com/file/d/1_z9w9mkLTbnGG2U1Iz7urQMppiWP_Dmm/view?usp=sharing}{link} \\ 
\texttt{FGBrpRg0.75} & 10 & 0.85   & 0.186 & 4.0  & 0.75  & 0.214   & 17.1  & 0.883 & 3300 & \href{https://drive.google.com/file/d/1cRhIXrlComFRMUpcf9_vngk0qX-dvg9L/view?usp=sharing}{link} \\
\enddata
\end{deluxetable*}

\section{Hydrodynamics of black hole+giant collisions}
\label{sec:hydro}

In this section, we investigate the hydrodynamic outcome of black hole+giant collisions to explore under what circumstances these may successfully form a compact black hole+white dwarf binary that can enter Roche contact.

\begin{figure*}
    \centering
    \includegraphics[width=1\linewidth]{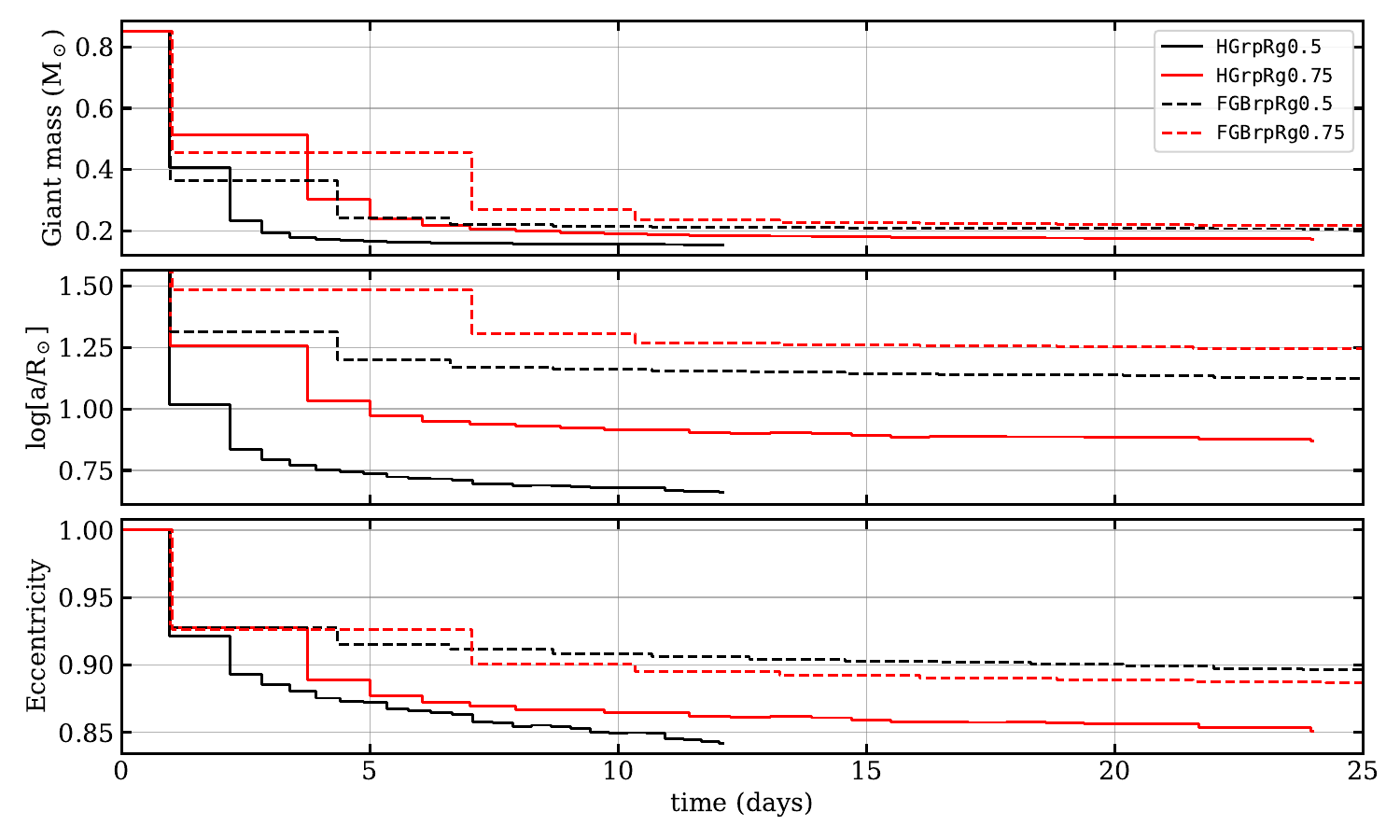}
    \caption{Properties of the binary during the SPH collisions between a $0.85\,M_\odot$ giant and a $10\,M_\odot$ black hole (see Table~\ref{table:SPH}). We track the giant's mass, semi-major axis, and eccentricity of the orbit versus time. The black hole is initially on a parabolic trajectory. At each periapsis passage, the orbital properties are updated to the values measured near the midpoint between that passage and the next.\label{fig:hydro}}
\end{figure*}


We use the smoothed-particle hydrodynamics (SPH) code \texttt{StarSmasher} \citep{gaburov2010} to compute the outcome of these collisions.
Each simulation employs approximately $10^5$ SPH particles.
Following recent work with \texttt{StarSmasher} \citep[e.g.,][]{Gibson2025,GonzalezPrieto2026}, we utilize the stellar evolution models from \texttt{MESA} \citep{paxton2011,paxton2013,paxton2015,paxton2018,paxton2019} for the initial stellar density profiles for the giants. Following \citet{kremer2022} and \citet{Lombardi2006}, the black holes and giant cores in our SPH models are
treated as point particles with a softened potential and
interact only via gravity with the gas SPH particles. 

We perform simulations assuming an initial black hole mass of $10\,M_{\odot}$ and giant mass of $0.85\,M_{\odot}$, motivated by the typical values of collisions we identify in our \texttt{CMC} models (Figure~\ref{fig:BHgiantmass}). We consider two different evolutionary times for the giant: (1) a Hertzprung gap case with a core mass of $0.14\,M_{\odot}$ and (2) a first-giant branch case with a core mass of $0.19\,M_{\odot}$. These are the two most common giant types identified in our \texttt{CMC} models. Additionally, we consider two values for the distance at closest approach of the initially parabolic encounter, $r_{\rm p}/R_{\rm G}=[0.5,0.75]$, to explore how the penetration factor of the collision impacts the final orbital properties of the binary formed. This gives us four SPH models in total. The motivation here is to identify outcomes for most typical cases. Of course, a broader suite of future models may explore a much more expansive set of parameters. We list the initial and final properties of our four models in Table~\ref{table:SPH}.

In Figure~\ref{fig:hydro}, we show the results for our four SPH simulations. The top panel shows the total mass of the giant as it is stripped, the middle panel shows the semi-major axis of the black hole+core binary, and the lower panel shows the orbital eccentricity. We compute the bound mass of the binary components using the iterative energy-based procedure of \citet{kremer2022}, and we determine the semimajor axis and eccentricity of the orbit from the Keplerian two-body orbital energy and angular momentum evaluated near apoapsis.
 In general, the giant rapidly sheds its outer layers, ultimately leaving behind just the low-mass core with perhaps a small remnant of the initial envelope (compare the initial core mass to final giant mass in Table~\ref{table:SPH}). We find that these collisions all produce binaries with high eccentricities and semi-major axes on the order of $10\,R_\odot$ or less, with the initial pericenter distance being the most significant factor that affects the final semi-major axis. Note that the high eccentricities computed here are consistent with the findings of \citet{ivanova2005} and \citet{Lombardi2006}, which both performed similar SPH simulations of neutron star+giant collisions.

Given the final masses and orbital properties at the end of our SPH simulations, we can compute the gravitational wave inspiral time using the equations of \citet{Peters1964}. We print the inspiral times in the final column of Table~\ref{table:SPH}. Consistent with the findings of \citet{ivanova2010}, the key takeaway is that these collisions lead to short inspiral times of order Gyr or less. Therefore, we can assume that the majority of these should successfully be able to enter Roche contact and form a UCXB by the present-day age of their host cluster. Extrapolating from the results here, we anticipate that more distant grazing encounters with $r_{\rm p}\gtrsim R_{\rm G}$ would lead to wider orbits that may not have time to enter Roche contact. 

Our SPH models carry several caveats worth noting. We adopt a gravitational softening length of $h = 0.16\,R_\odot$ for the black hole, consistent with the approach of \citet{kremer2022}. We also adopt an accretion radius of 74 km: SPH particles that pass within this distance from the black hole are removed from the simulation, with their mass and momentum added to the black hole particle. Although a small accretion disk visibly forms around the black hole in each simulation, we do not attempt to model the evolution of this disk or any associated feedback. Additionally, we do not account for recombination energy, which has been shown to be an important energy source in common envelope events \citep[e.g.,][]{ivanova2013,nandez2015}. These simplifications are unlikely to alter our primary conclusion that black hole+giant collisions produce compact, highly eccentric binaries with short gravitational wave inspiral timescales. A more detailed treatment of the microphysics is left to future work.

\section{Modeling stable mass transfer and X-ray luminosities}
\label{sec:MT_Xray}

Once the black hole+white dwarf binary ultimately comes into Roche contact via gravitational wave emission, we use the binary evolution code \texttt{COSMIC} to model the subsequent mass transfer \citep{breivik2020}. \texttt{COSMIC} employs the mass transfer physics outlined in \citet{Hurley2002}. Of course, a population synthesis code like \texttt{COSMIC} represents a simplified approach to this problem compared to more sophisticated approaches like those employed in \citet{Church2017}. However, the goal here is an order-of-magnitude-style calculation, for which the \citet{Hurley2002} approach is sufficient.

We calculate the accretion luminosity from the mass transfer rate $\dot M$ using the standard expression \mbox{$L=\eta \dot M c^2$}, where $\eta$ is the radiative efficiency. We assume $\eta=0.1$ for black holes accreting in the soft state and that $\eta \propto \dot M$ in the hard state and quiescence. We also assume a continuous transition between the two regimes and that systems transition from hard to soft states at a critical accretion rate $\dot{M}_{\rm tr}=0.02\,M_{\rm Edd}$ or roughly $2.5\times 10^{37}\,\rm{erg\,s}^{-1}$ for a $10\,M_\odot$ black hole \citep[for further discussion, see][]{Maccarone2003,NarayanMcClintock2008,Tudor2018,bahramian2023}. The (bolometric) accretion luminosity is then given by:

\begin{equation}
    L =
    \left\{
    \begin{array}{@{}l@{\quad}l@{\quad}r@{}}
        \eta_0 \dot{M}c^2
        & \text{if } \dot{M} > \dot{M}_{\rm tr} = 0.02\dot{M}_{\rm Edd}
        & \text{(soft state)}
        \\[4pt]
        \eta_0 \dfrac{\dot{M}^2}{\dot{M}_{\rm tr}}c^2
        & \text{if } \dot{M} \leq \dot{M}_{\rm tr}
        & \text{(hard state)}
    \end{array}
    \right.
    \label{eq:Lx}
\end{equation}
where $\eta_0=0.1$. For simplicity, we assume the majority of this power will be emitted in the X-ray spectral range, so for our purposes, this expression is equivalent to the X-ray luminosity.

\begin{figure}
    \centering
    \includegraphics[width=\linewidth]{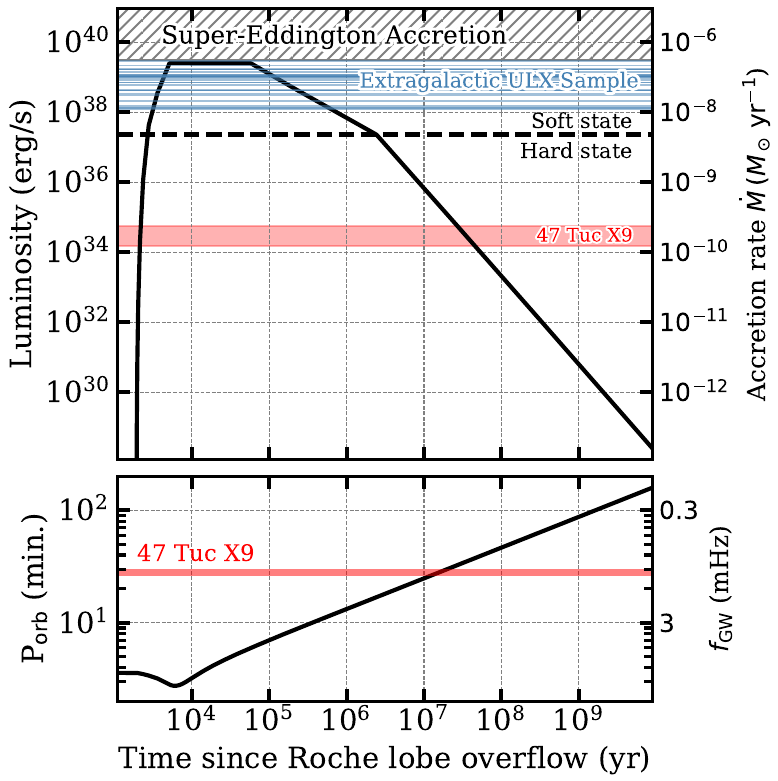}
    \caption{\textit{Top panel:} Black curve shows our computed X-ray luminosity versus time for a fiducial UCXB consisting of a $10\,M_\odot$ black hole accretor and a $0.2\,M_{\odot}$ helium white dwarf donor. Luminosity is computed using Equation~\ref{eq:Lx}, where the horizontal dashed line denotes our assumed value for $\dot{M}_{\rm tr}$ where the source transitions from soft to hard state. The hatched gray region denotes luminosities/accretion rates corresponding to super-Eddington accretion for our assumed black hole mass. Blue bands show the combined extragalactic globular cluster ULX sample from \citet{Dage2019,dage2020,dage2021}. We also show as a red band the range in X-ray luminosity for the 47~Tuc~X9 Galactic UCXB candidate \citep{Tudor2018}. \textit{Bottom panel:} Orbital period and corresponding gravitational wave frequency versus time. As the system undergoes stable mass transfer onto a higher-mass accretor, the orbital separation increases, driving the system to higher orbital periods (lower gravitational wave frequencies).\label{fig:Mdot}}
\end{figure}

In Figure~\ref{fig:Mdot}, we show luminosity versus time (top panel) from the onset of Roche lobe overflow for a fiducial UCXB with $M_{\rm BH} = 10\,M_\odot$ and $M_{\rm WD} = 0.2\,M_\odot$ (see Figure~\ref{fig:BHgiantmass}). The orbital evolution and corresponding mass transfer rate (approximate $\dot{M}$ values are shown on the secondary y-axis for the top panel) are computed using \texttt{COSMIC} and luminosity is computed using Equation~\ref{eq:Lx}. The hatched gray region at the top of the figure denotes the Eddington-limited accretion rate, which as shown, is assumed as a hard limit in the maximum accretion rate of the black hole. In the bottom panel of the figure, we show the binary's orbital period and corresponding gravitational wave frequency versus time, defined as $f_{\rm GW} = 2 f_{\rm orb}$, where $f_{\rm orb}$ is the system's orbital frequency (assuming a circular orbit). Because mass transfer occurs from a lower-mass donor, mass transfer causes the system to expand. Note this expansion is in opposition to gravitational wave emission which acts to drive the system to small separations \citep[higher frequencies; for discussion of similar sources, see e.g.,][]{tauris2018}.

As shown, a typical UCXB will quickly evolve through the initial phase of high accretion rates. Assuming a luminosity threshold of roughly $10^{39}\,\rm{erg\,s}^{-1}$ for ultraluminous X-ray sources (ULXs), the ULX lifetime is roughly $\tau_{\rm ULX}\approx 10^5\,$yr. For reference, the horizontal blue bands in the figure mark the observed luminosities for extragalactic ULXs from \citet{Dage2019,dage2020,dage2021}. On slightly longer timescales, the source transitions from soft to hard state (shown as dashed black line in the figure) with an approximate lifetime for soft-state accretion of $\tau_{\rm soft} \approx 10^6\,$yr. Assuming a formation rate per cluster of $\mathcal{R}\approx 6\times 10^{-10}\,\rm{yr}^{-1}$ (based on our \texttt{CMC} models described in Section~\ref{sec:cluster_properties}), and assuming a population of $10^4$ globular clusters \citep[e.g., representative of the population of clusters in Virgo;][]{Jordan2009}, we can then compute a rough estimate for the number of observable ULXs formed via black hole+giant collisons as
\begin{equation}
    \label{eq:ULX}
    N_{\rm ULX} \approx 0.5\Bigg(\frac{\mathcal{R}}{6\times10^{-10}\,\rm{yr}^{-1}} \Bigg) \Bigg(\frac{N_{\rm GC}}{10^4} \Bigg) \Bigg(\frac{\tau_{\rm ULX}}{10^5\,\rm{yr}} \Bigg).
\end{equation}

On longer timescales, the source is no longer bright enough to be visible in extragalactic globular clusters, but still could be observed as a quiescent source in the Milky Way, similar to 47~Tuc~X9 \citep[the estimated bolometric luminosity is marked as a red band in the figure;][]{Tudor2018}. Based on Figure~\ref{fig:Mdot}, we can estimate a typical lifetime for a 47~Tuc~X9-like source to be $\tau_{\rm X9} \approx 3\times10^7\,$yr. Performing a similar estimate for the number of Galactic sources, we predict
\begin{equation}
    \label{eq:X9}
    N_{\rm X9} \approx 2\Bigg(\frac{\mathcal{R}}{6\times10^{-10}\,\rm{yr}^{-1}} \Bigg) \Bigg(\frac{N_{\rm GC}}{150} \Bigg) \Bigg(\frac{\tau_{\rm X9}}{3\times 10^7\,\rm{yr}} \Bigg).
\end{equation}

Of course, these are rough estimates that do not take into account the diversity of sources formed (for example, more massive black holes produce brighter luminosities) nor the diversity in formation rate across different cluster properties (Section~\ref{sec:cluster_properties}). Nonetheless, it is reassuring that these order-of-magnitude estimates are consistent to within a small factor of the observed number of sources in both the Milky Way and in extragalactic clusters. These rates are also consistent with similar estimates in the literature, namely \citet{Church2017} (specifically, see their Figure 3 which is closely related to our Figure~\ref{fig:Mdot}) as well as \citet{Tudor2018}. These studies both highlight the X-ray source RZ 2109 \citep[localized to a globular cluster in Virgo;][]{Maccarone2007} as a possible younger and brighter version of 47~Tuc~X9, and estimate similar Galactic and extragalactic source counts for these types of systems.

We also note that this calculation predicts a much larger sample of lower-luminosity sources in the Milky Way globular clusters. Our model predicts sources with luminosities of roughly $10^{30}\rm{erg\,s}^{-1}$ to have characteristic lifetimes of roughly $10^9\,$yr. From Equation~\ref{eq:X9}, this would correspond to dozens of additional low-luminosity black hole+white dwarf binaries in the Milky Way.

\section{Building a mock sample of the local ultracompact binary population}
\label{sec:mocksample}

Motivated by the rough estimates from Equations~(\ref{eq:ULX}) and (\ref{eq:X9}), we now perform a more detailed calculation of the number of observable UCXBs formed via black hole+giant collisions. We synthesize mock samples of black hole+white dwarf X-ray binaries using a Monte Carlo method by performing random draws of the \texttt{CMC} N-body models weighted by properties of observed globular clusters in the Milky Way and Virgo clusters, then identifying the luminosities of relevant black hole+white dwarf systems within each model. 

\subsection{Observation to model assignment}

\begin{figure*}
    \centering
    \includegraphics[width=\linewidth]{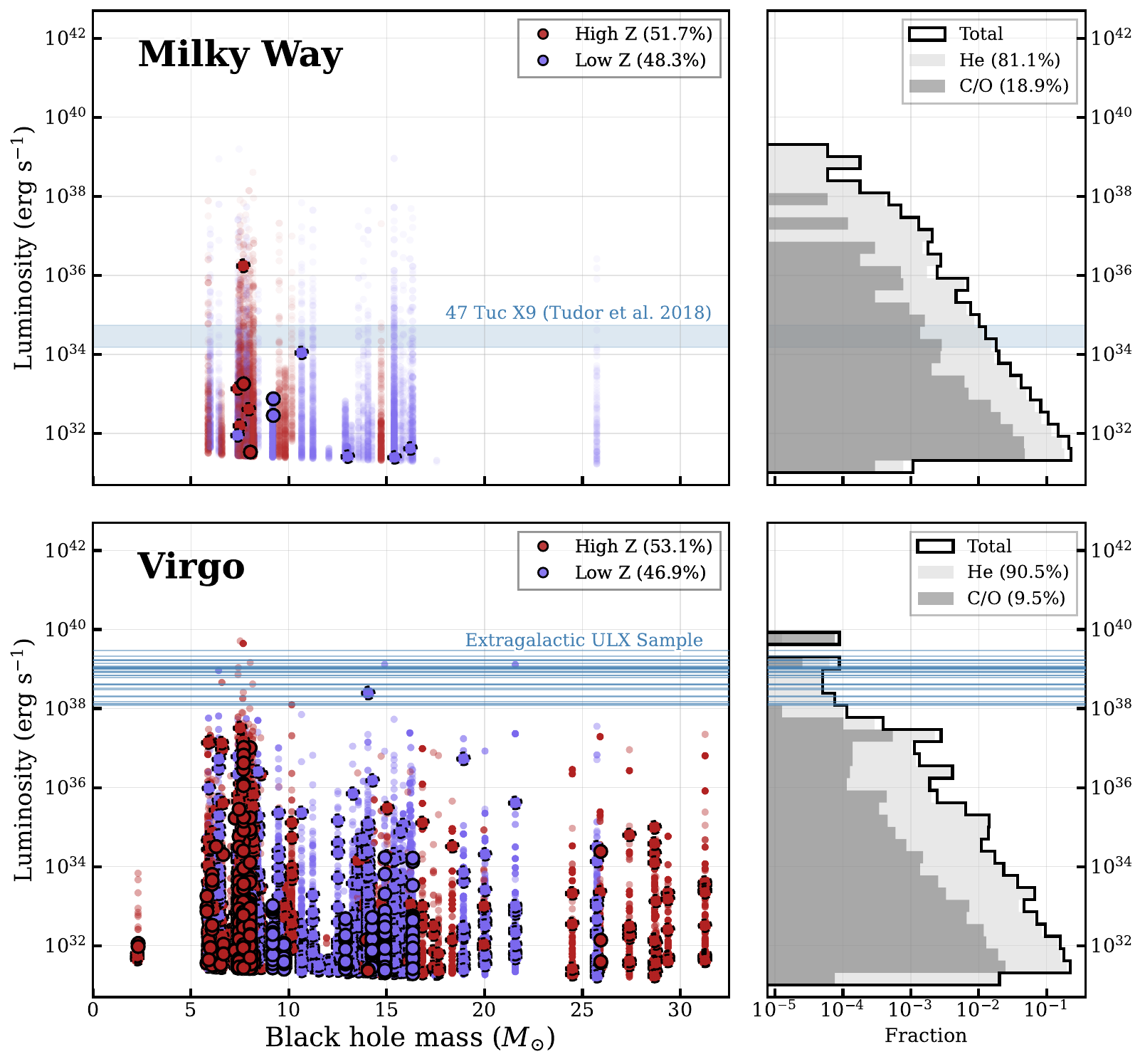}
    \caption{Generated sample of UCXBs in the globular clusters in the Milky Way (\textit{top panels}) and Virgo (\textit{lower panels}), following the method outlined in Section~\ref{sec:mocksample}. The systems are divided into high metallicity and low metallicity bins with a cutoff at $\text{Fe}/\text{H}=-0.75$ ($Z=0.18 Z_\odot$). Scatter points with a solid outline contain a C/O white dwarf, while dashed points contain a He white dwarf. Histograms displaying the distribution of C/O, He, and all systems in all Monte Carlo draws are shown on right. \label{fig:samples}}
\end{figure*}

We first construct a model grid using a set of late time snapshots of the \texttt{CMC} simulations. We then match observational catalogs of the Milky Way and the Virgo Cluster globular clusters from \citet{Harris1996,Jordan2009} against this grid using a procedure that largely follows \citet{Mai2026}.

To summarize the key steps, a unique best matching model is assigned following a weighted least squares procedure, minimizing the distance function
\begin{equation}
    d^2=(C\cdot\Delta \log M_{\rm cl})^2 + (\Delta r_h)^2
\end{equation}
with the constant $C=10$ to balance the scales of $\log M_{\rm cl}$ and half-light radius $r_h$. 

As we only have a discrete grid of \texttt{CMC} models, we perform a rescaling procedure to the expected number of black hole+giant collisions from the matched model to that of the observed cluster using the relations outlined in Section~\ref{sec:cluster_properties}. We use the approximate result
\begin{equation}
    N_{\rm BHG} \propto M_{\rm cl} r_h^{-2}
\end{equation}
to produce an interpolating factor
\begin{equation}
    f_{\rm interp} = \frac{M_{\rm cl, obs}}{M_{\rm cl,CMC}} \left(\frac{r_{h,\rm obs}}{r_{h,\rm CMC}}\right)^{-2}.
\end{equation}
The expected number of collisions assigned to the observed cluster is then
\begin{equation}
    \langle N_{\rm coll,obs} \rangle = f_{\rm interp} N_{\rm coll,CMC}.
\end{equation}

Motivated by the strong evidence for significant relationships between event rate and cluster metallicity in Section~\ref{sec:cluster_properties}, each observed cluster is divided into two metallicity bins intended to correspond to the ``red'' and ``blue'' globular cluster subpopulations demarcated at the boundary [Fe/H]$=-0.75$ \citep[e.g.,][]{BrodieStrader2006}. Observed values for Milky Way clusters are obtained directly from \citet{Harris1996}. For Virgo clusters, we assume a 2:1 ratio of blue to red globular clusters similar to what is observed for M87 which contains the majority of the globular clusters in Virgo \citep[e.g.,][]{Harris2009,Strader2011,Villaume2019}. For our
\texttt{CMC} models, we group our high metallicity ($Z=Z_\odot$) models into the red cluster population, and our low metallicity models, $Z= [0.01,0.1]Z_\odot$, into the blue population.

We sample globular cluster ages by randomly sampling from the range $9-13.7\,$Gyr for the assigned \texttt{CMC} model. For each observed cluster we perform $\langle N_{\rm coll,obs} \rangle$ age draws, corresponding to the expected number of UCXB-formation events (black hole+giant collisions) in that cluster. Each late-time black hole+giant collision product is then evolved forward using \texttt{COSMIC} for the remaining time between the collision event and the randomly drawn cluster age, following the method in Section~\ref{sec:MT_Xray}. From this evolution, we compute the accretion luminosity for all resulting black hole+white dwarf binaries expected in the cluster at present. Because we are sampling from a finite set of \texttt{CMC} models (and therefore a finite set of UCXBs), we inevitably sample each black hole+giant collision pair multiple times.

Finally, this process is repeated a number of times corresponding to the number of globular clusters of interest: $150$ for the Milky Way, and $20,000$ for the Virgo Cluster. This yields Figure~\ref{fig:samples}, which shows computed X-ray luminosity versus black hole mass for all UCXBs predicted in the globular clusters within the Milky Way and Virgo. Each panel oversamples the number of sources to better resolve the statistical sample: for the Milky Way, we oversample by a factor of 100 and for Virgo, by a factor of 10. The smaller transparent scatter points show our full (oversampled) population and the solid larger points show the results for a single representative realization for each group. As in Figure~\ref{fig:Mdot}, we mark the observed X-ray luminosities for 47~Tuc~X9 and the ULX sources from \citet{Dage2019,dage2020,dage2021}.



Scatter points with (without) a black border denote systems with a C/O (He) white dwarf donor. We also separate these two donor types in the histograms shown to the right. As shown, the vast majority (roughly $80-90\%$) of our UCXBs feature He white dwarf donors. As discussed earlier in the context of Figure~\ref{fig:BHgiantmass}, this is because giants spend less time during their core He burning phase and therefore are less likely to experience a black hole collision that would result in a C/O white dwarf. 

Red and blue colors in Figure~\ref{fig:samples} denote the metallicity of the host cluster in which each UCXB originates, with red (blue) denoting the high (low) metallicity globular clusters as defined above. We find that over half of our UCXBs are formed in high metallicity red clusters, notable given that red clusters constitute only a third of the cluster population in Virgo and less than a third in the Milky Way. This enhancement arises from the strong preference for black hole+giant collisions in high metallicity clusters, shown in Figures~\ref{fig:correlation_N} and \ref{fig:correlation_rv}. In general, the higher metallicity clusters form lower-mass black holes and vice versa, a consequence of our metallicity-dependent stellar wind prescriptions (see Section~\ref{sec:methods}. The small number of high-metallicity (red) black holes in Figure~\ref{fig:samples} with masses above $25\,M_{\odot}$ are grown through previous binary black hole mergers and/or stellar collisions.

\subsection{Comparison to Observed Candidates}
\label{sec:observed_UCXB}


\begin{figure*}[ht!]
    \centering
    \includegraphics[width=0.8\linewidth]{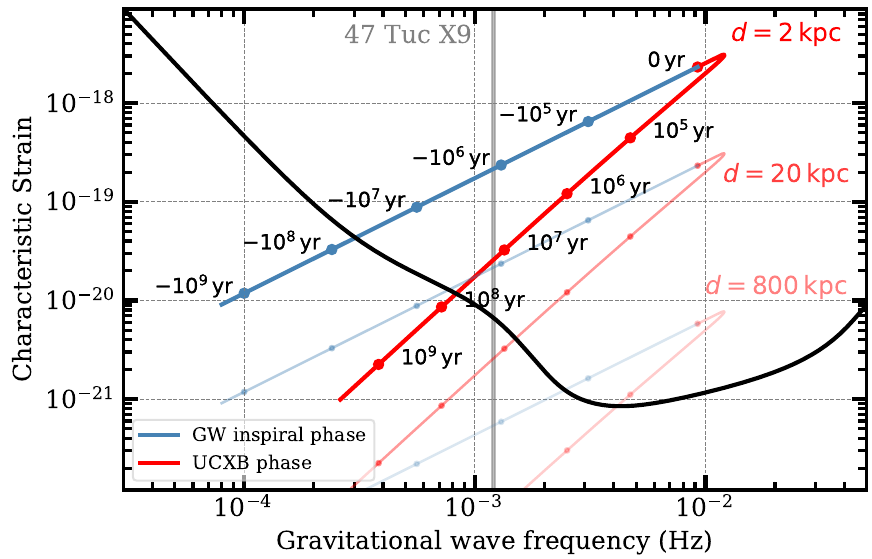}
    \caption{LISA characteristic strain amplitude vs gravitational wave frequency for our fiducial source consisting of a $0.2\,M_\odot$ white dwarf and a $10\,M_\odot$ black hole (see Figures~\ref{fig:BHgiantmass} and \ref{fig:Mdot}). Blue segments of each curve denote the GW-driven inspiral phase and red segments denote the ultracompact X-ray binary phase. The evolution rate has been indicated with time intervals on each track, where $t=0$ has been defined to be at mass transfer turn-on. We show curves for three separate distances, $[2,20,800]\,$kpc, respectively corresponding roughly to the closest Galactic globular clusters, the furthest Galactic clusters, and clusters in M31. The vertical gray line denotes the inferred gravitational wave frequency for the black hole candidate 47~Tuc~X9 \citep{bahramian2017}. The solid black curve shows the LISA sensitivity curve.}
    \label{fig:gwstrain}
\end{figure*}

\begin{figure*}[ht!]
    \centering
    \includegraphics[width=1.0\linewidth]{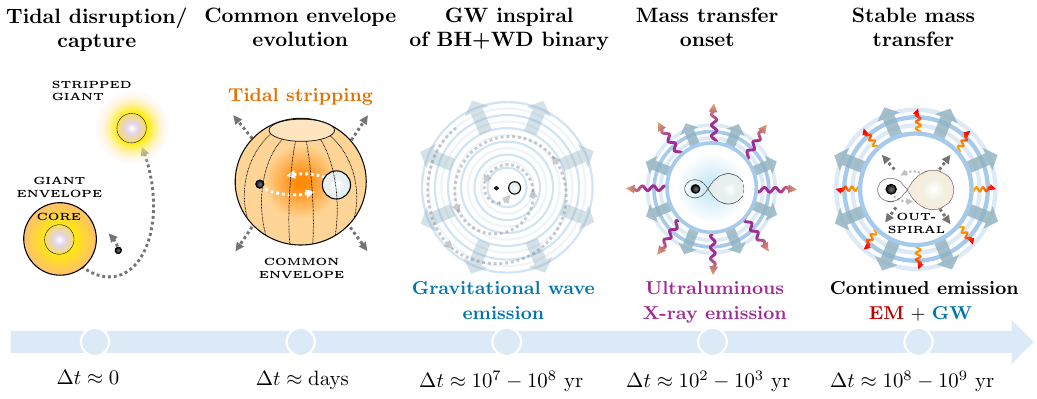}
    \caption{Schematic illustration of the formation of an ultracompact black hole X-ray binary via a black hole+giant collision. From left to right, we show: (1) A black hole+giant collision that unbinds the outer envelope of the giant, with the giant core tidally captured into an elliptical orbit with the black hole; (2) Formation of a common envelope around the black hole+core binary system which is stripped via tidal ejection and/or accretion into the black hole; (3) The black hole+core binary evolves into a black hole+white dwarf binary and inspirals from gravitational wave emission; (4) Roche lobe overflow leads to stable mass transfer initially as an ultraluminous X-ray source with $L\gtrsim 10^{38}\,\mathrm{erg\,s}^{-1}$ \citep[similar to known extragalactic ULXs; e.g.,][]{Dage2019,dage2020,dage2021} for roughly $10^5\,$yr; and (5) on longer timescales (roughly $10^7-10^8\,$yr), a lower-luminosity ($L\approx10^{32}-10^{36}\,\mathrm{erg\,s^{-1}}$) X-ray source similar to Galactic sources like 47~Tuc~X9. During both stages (4) and (5), the accreting binary is simultaneously a source of millihertz gravitational wave emission.\label{fig:schematic}}
\end{figure*}

Up to $20$ ULXs are known at present that have been localized to extragalactic globular clusters \citep[e.g.,][]{Maccarone2007,Irwin2010,Dage2019,dage2020,dage2021}. This is roughly a factor of ten higher than the number of ULXs we predict in Figure~\ref{fig:samples}. However, not all observed ULXs are necessarily ultracompact binaries with degenerate donors. Spectral features may shed light on the nature of the companion. For example, optical spectroscopy of RZ~2109 in NGC~4472 reveals very broad and luminous [OIII] emission lines, while also having no detectable hydrogen emission \citep{Zepf2007,Zepf2008,Steele2014,Dage2019b}, potentially hinting at a degenerate donor. The ULX source GCU7 \citep[located in NGC~1399;][]{Irwin2010} also exhibits [OIII] emission, and thus has also been touted as another strong UCXB candidate \citep{Oh2025}. Note that GCU7 has similar inferred X-ray luminosity to RZ 2109 \citep[e.g.,][]{Dage2019}, so in our mass transfer model, we expect these two sources to have roughly similar UCXB ages. Meanwhile, no evidence of [OIII] optical emission was found for the ULX GCU1, also in NGC~4472 \citep{Athukoralalage2023}. Future optical spectroscopy of other ULX candidates is needed to determine whether [OIII] emission is rare or a common feature of these sources. Therefore, for the full sample of roughly 20 known globular cluster ULXs, only two could be considered bona fide UCXB candidates at present. Our predicted value of roughly one ULX source in Virgo shown in Figure~\ref{fig:samples} is consistent with this value. 

The donor star for 47~Tuc~X9 is widely interpreted to be a C/O white dwarf, as a result of strong carbon lines in the UV spectrum \citep{Knigge2008}, absence of helium lines, and evidence of oxygen abundance in the photoionized gas from X-ray spectral fitting \citep{bahramian2017}. From the Monte Carlo population synthesis, we predict roughly $1.2 \pm 1$ UCXBs sources with luminosities of $10^{34}\,\rm{erg\,s}^{-1}$ or more in the Milky Way, with the $90^{\rm th}$ percentile boundary at 3 UCXB sources. For our Galactic sample, roughly $20\%$ of sources feature a C/O white dwarf donor (see Figure~\ref{fig:samples}). In this case, our model is consistent to within a small factor with the presence of a single X9-like source.

Order hundreds of lower-luminosity X-ray binaries are also observed in association with extragalactic globular clusters out to distances of Virgo and Fornax down to limiting
magnitudes of roughly $10^{37}\,\rm{erg\,s}^{-1}$ \citep[e.g.,][]{Sarazin2001,Kundu2003,Irwin2005,VerbuntLewin2006,Kim2006,Kundu2007,Kim2013,Peacock2017,Lehmer2020}. In our mock Virgo population, we observe $32\pm6$ sources at present with luminosities at or above this threshold, ranging from $9$ to $146$ across the Poisson distributed realizations, with $1.4\pm1$ ultraluminous sources ($\gtrsim 10^{39}\,{\rm erg/s}$). This is reasonable, as we expect UCXB with black hole accretors to constitute just a fraction of the full X-ray binary population, the majority of which presumably are neutron star systems \citep[e.g.,][]{VerbuntLewin2006}.

A notable result for the full sample of X-ray sources is a clear link between optical color/metallicity of their host globular cluster. \citet{Kundu2003} shows that metal-rich clusters are roughly three times more likely to host regular low mass X-ray binaries. This is notable, as metal-rich ``red'' globular clusters typically constitute only a third of the full globular cluster population \citep[e.g.,][]{BrodieStrader2006}, hinting that the enhancement of X-ray binaries in metal-rich clusters may be up to a factor of ten compared to lower-metallicity systems. We find a similar preference for UCXBs to form in metal-rich clusters (see Figures~\ref{fig:correlation_N} and \ref{fig:correlation_rv}). In our mock catalog of Virgo sources, we identify a global enhancement of roughly 1.2 in metal-rich clusters, compared to a factor of three in \citet{Kundu2003}. Again, we emphasize that UCXBs formed via black hole+giant collisions are just one pathway for forming low-mass X-ray binaries in clusters, and presumably the majority of lower-luminosity X-ray binaries are simply neutron star binaries \citep[analogous to the type of systems that form the abundant population of millisecond pulsars in globular clusters; e.g.,][]{Ransom2008}. \citet{Nair2023} examined whether the subpopulation of ULXs exhibit a similar preference for metal-rich hosts, and did not find a clear trend. Further work is necessary to explore these potential correlations for the complete observed UCXB sample. 

\vspace{1cm}

\section{Gravitational Wave Signal}
\label{sec:GWs}

As a final step, we explore the detectability of these UCXBs as millihertz gravitational wave sources for LISA. In principle, these sources could be observable during their initial inspiral phase (following common envelope and prior to the onset of the UCXB phase) and during the UCXB phase itself when the binaries orbit is expected to outspiral in response to mass transfer (see Figure~\ref{fig:Mdot}). We explore both possibilities here.

We use the expressions defined by \citet{tauris2018} to calculate the characteristic strain observed during the LISA mission,
\begin{equation}
    \label{eq:GWstrain}
    h_c\approx\sqrt{N_{\rm cycles}}\sqrt{2}h_0=\sqrt{2f_{\rm gw}T}h_0
\end{equation}
where $N_{\rm cycles}$ is the number of cycles for an observation time $T=4\,$yr, $f_{\rm GW}$ is the gravitational wave frequency $f_{\rm GW}$, and where
\begin{equation}
    h_0=\left(\frac{32}{5}\right)^\frac{1}{2}\frac{\pi^{2/3}G^{5/3}f_{\rm gw}^{2/3}\mathcal{M}^{5/3}}{c^4d_L}
\end{equation}
is the gravitational wave amplitude for a binary at a luminosity distance $d_L$, and $c$ and $G$ are their respective physical constants. Our binary evolution models performed with \texttt{COSMIC} track the binary separation, masses, and time scales on which the systems evolve, which can then be plugged into Equation~\ref{eq:GWstrain} to compute the gravitational wave amplitude. 

In Figure~\ref{fig:gwstrain}, we plot the characteristic strain versus gravitational wave frequency for our fiducial source with $M_{\rm BH}=10\,M_\odot$ and $M_{\rm WD}=0.2\,M_\odot$ also shown in Figure~\ref{fig:Mdot}. We adopt three characteristic distances: $2\,$kpc \citep[corresponding to the closest Galactic globular clusters;][]{Harris1996}, $20\,$kpc (corresponding to roughly the $80^{\rm th}$ percentile of Galactic clusters), and $800\,$kpc (corresponding to globular clusters in M31). Blue portions of the curves show the pure gravitational-wave inspiral phase prior to Roche contact, and red portions correspond to the UCXB phase shown in Figure~\ref{fig:Mdot}. We mark with scatter points key timescales before and after the onset of mass transfer. As shown, the binary's characteristic strain increases as the system inspirals up to Roche contact, and then begins to decrease once mass transfer begins. The decrease in strain is driven by both the increasing frequency as well as the decreasing chirp mass as the system undergoes mass transfer; as a result, the strain decreases rapidly once the UCXB phase begins. For comparison and further discussion, see \citet{tauris2018} who computes analogous curves for UCXBs consisting of neutron star accretors.

The signal-to-noise ratio of the source can be approximated ``by eye'' as the height of the characteristic strain above the LISA noise curve \citep[shown as the solid black curve in the figure, and adopted from][]{Robson2019}. As shown, a UCXB could be readily resolved as a LISA source in one of the closest Galactic clusters (e.g., M4 or NGC~6397) on timescales of roughly $10^7\,$yr post-Roche contact (as well as roughly $10^7\,$yr pre-Roche contact). However, given our predicted formation rate ($\mathcal{R}\approx 6\times10^{-10}\,\rm{yr}^{-1}$), the chances of such a source being present are slim. Out to largest distances that encompass the full population of roughly $150$ Galactic clusters, we predict order unity sources may potentially be observable as LISA sources. Indeed, 47~Tuc~X9 is a potential case-in-point; we show as the vertical gray line the inferred gravitational wave frequency of this source which is firmly within LISA's sensitivity range. Based on its current orbital period and accretion history (adopting the fiducial model of Figure~\ref{fig:Mdot}), we predict 47~Tuc~X9 has a current companion mass of roughly $10^{-2}\,M_{\odot}$, consistent with estimates of \citet{bahramian2017,Church2017,Tudor2018}. Adopting a black hole mass of $10\,M_{\odot}$ and distance to 47~Tuc of roughly $4.5\,$kpc \citep{Harris1996}, we predict a signal-to-noise ratio of order $0.1$. Thus, as also predicted in \citet{Tudor2018}, we find 47~Tuc~X9 is just below the LISA detection threshold. However, an analogous UCXB that is slightly younger (closer to the onset of mass transfer) or in a slightly closer globular cluster could potentially be detectable.

Last, we note that extragalactic UCXBs are likely not viable as LISA sources. As shown in the figure, a source in M31 could optimistically be detected for roughly $10^4\,$yr post-Roche contact. M31 has roughly $500$ globular clusters \citep[e.g.,][]{Harris2013}. Adopting our same formation rate per cluster as before, this suggests the chances of such a source being observed in M31 are less than $1\%$. Sources in more distant galaxies (e.g., in Virgo) are too distant to be resolved at any point of their evolution.

\section{Conclusions and discussion}
\label{sec:conclusions}

\subsection{Summary}

To summarize, we illustrate in Figure~\ref{fig:schematic} the key evolutionary phases considered, including the initial black hole+giant collision (Section~\ref{sec:cluster_properties}), inspiral of the black hole+core during common envelope (Section~\ref{sec:hydro}), inspiral via gravitational wave emission of the black hole+white dwarf binary, and ultimate onset of stable mass transfer and X-ray emission (Section~\ref{sec:MT_Xray}).

Our key findings are:

\begin{enumerate}
    \item Using the $N$-body models from the \texttt{CMC Cluster Catalog} \citep{kremer2020b}, we explore several pathways through which ultracompact X-ray binaries consisting of black hole accretors and white dwarf donors may form. Although many interacting black hole+white dwarf binaries can form via exchange encounters, the high white dwarf donor masses (typically $1\,M_{\odot}$ or more) for these systems lead to unstable mass transfer and merger, rather than a long-lived X-ray source. This is a key difference from the conclusions of previous work that used semi-analytic calculations to study exchange interactions \citep[e.g.,][]{ivanova2005,ivanova2010}.
    \item Instead, our $N$-body models show that black hole+giant collisions are the dominant mechanism through which ultracompact black hole X-ray binaries form. Using our \texttt{CMC} models, we estimate a black hole+giant collision rate of roughly $6\times10^{-10}\,\rm{yr}^{-1}$ per cluster, on average. We show this rate can vary by up to several orders of magnitude for various cluster properties, namely cluster mass $M_{\rm cl}$ and virial radius $r_v$. We find the total number of black hole+giant collisions can be approximated using the functional form $N_{\rm BHG} \propto M_{\rm cl} r_v^{-2}$.
    \item Notably, we find the black hole+giant collision rate is enhanced by a factor of roughly ten in our highest-metallicity \texttt{CMC} models. This is because higher metallicity models generally feature lower-mass black holes, which are able to dynamically mix with luminous stars more effectively \citep[see also,][]{kremer2020b}.
    \item We perform a set of four hydrodynamic simulations of typical black hole+giant collisions using \texttt{StarSmasher}. In all cases, we find the giant envelope is nearly entirely expelled, leaving behind a compact ($a\lesssim 10\,R_{\odot}$) and eccentric ($e\gtrsim0.8$) binary. These binaries have gravitational wave inspiral times of order Gyr or less, indicting they will successfully enter Roche contact within a Hubble time, and form a UCXB consisting of a black hole accretor and low-mass white dwarf donor.
    \item By modeling the subsequent stable mass transfer, we find the UCXB spends roughly $10^6\,$yr as an ultraluminous X-ray source with $L\gtrsim10^{38}\,\rm{erg\,s}^{-1}$, analogous to UCXB candidate sources that have been identified in extragalactic globular clusters \citep{Maccarone2007,Dage2019,dage2020,dage2021}. On longer timescales ($\gtrsim 10^7\,$yr), the mass transfer rate drops and the UCXB transitions to a hard state. This second phase may yield sources similar to 47~Tuc~X9 in the Galactic cluster 47~Tuc \citep{bahramian2017}.

    \item On even longer timescales ($\gtrsim 10^9\,$yr), these sources may be detectable as much lower luminosity sources, down to roughly $10^{30}\,\rm{erg\,s}^{-1}$. We predict dozens of these lower-luminosity black hole UCXBs should be present in Milky Way globular clusters today.
    
    \item Combining our full set of \texttt{CMC} models with our hydrodynamic simulations and mass transfer calculations, we build a mock sample of UCXBs expected in the globular clusters of both the Milky Way and nearby galaxies. For each set of Poisson distributed realizations, we predict $1.2 \pm 1$ UCXBs sources with luminosities comparable to 47 Tuc X9 ($\gtrsim 10^{34}\,\rm{erg\,s}^{-1}$) in Galactic clusters today, and $32\pm6$ luminous sources ($\gtrsim 10^{37}\,{\rm erg/s}$) and $1.4\pm1$ ultraluminous sources ($\gtrsim 10^{39}\,{\rm erg/s}$) in the wider population of roughly $10^4$ globular clusters in Virgo.
    
    \item Our models predict a clear preference for black hole UCXBs to be found in metal-rich globular clusters, as a consequence of  enhanced dynamical mixing between black holes and stars for lower-mass black hole populations. Our predicted enhancement is qualitatively consistent with the observed enhancement of X-ray sources in metal-rich clusters \citep[e.g.,][]{Kundu2003}.
    
    \item Finally, we compute the strength of the gravitational wave signal for these UCXBs both pre- and post-Roche contact. We predict a small number (order unity) of these sources may be resolvable by LISA within nearby Galactic clusters. With an orbital period of roughly 30 minutes, 47~Tuc~X9 is a strong candidate for such a source, although we find it is likely just below LISA's detection threshold, consistent with conclusion of \citet{Tudor2018}.
\end{enumerate}

\subsection{Future Work}

Our conclusions set the stage for a number of future extensions. We have focused here on black hole+giant collisions; however the majority of X-ray sources in globular clusters are likely comprised of neutron star accretors. This is evidenced in part by the significant number of millisecond pulsars observed in Milky Way clusters \citep[e.g.,][]{Ransom2008}, presumably formed via spin up in an X-ray binary \citep[e.g.,][]{PhinneyKulkarni1994, Ye2019_psr}. Previous work has shown that neutron star+giant collisions can provide a similar pathway for forming neutron star UCXBs \citep[e.g.,][]{ivanova2005}. Future work may investigate the complete population of UCXBs using $N$-body models similar to those studied here. This is especially relevant to investigate the enhancement of UCXBs with metallicity that we have identified in this study. This would also benefit from a finer sampling in cluster metallicity (here we have considered just three metallicity values that coarsely span the full range observed), to more thoroughly examine how the number of X-ray sources of all types varies accross all cluster metallicities.

We have explored here the evolution of UCXBs in our \texttt{CMC} models in ``post-processing.'' As a result, we have ignored the possible role of longer-term dynamics that may impact the subsequent evolution of a UCXB after formation. In principle, later dynamical encounters in the host cluster center may alter the orbital properties of a UCXB, or potentially even disrupt a UCXB entirely mid-mass transfer. Given that the orbital separations of UCXBs are quite small (less than $0.1\,R_{\odot}$ once mass transfer begins), the cross section for encounters is tiny and it is quite reasonable to assume that encounters are rare on short timescales. However, for longer-lived systems (for example, 47~Tuc~X9 with an estimated age of roughly $10^7-10^8\,$yr post-Roche lobe overflow onset), such encounters may become important. Future studies may implement formation and evolution of UCXBs into an N-body code like \texttt{CMC}, so that the subsequent mass transfer and dynamical evolution can be computed self-consistently.

The \texttt{CMC} models used in this study adopt a constant binary fraction of $5\%$ for our initial stellar population. This choice is motivated by low binary fractions observed for the majority of Galactic globular clusters at present \citep[e.g.,][]{Milone2012}. However, a number of studies have pointed out that the binary fraction for high mass stars in clusters may be significantly higher \citep[e.g.,][]{Leigh2015,Belloni2017,OConnor2026}. Higher binary fractions can significantly impact the dynamics of massive stars and the compact objects that are ultimately formed \citep[e.g.,][]{Gonzalez2021}. For our purposes, higher binary fractions for black hole and neutron star progenitors may significantly impact the rate of X-ray sources formed, both ultracompact and otherwise. Future work may explore a broader range of primordial binary fractions and properties, and how these uncertainties impact our conclusions.

Finally, a key conclusion here is that that black hole+white dwarf binaries formed via exchange interactions likely lead to unstable mass transfer and merger upon Roche contact, as a result of the high white dwarf masses (Figure~\ref{fig:bhwd_exchange}). Although these mergers would not lead to formation of a long-lived UCXB, they may produce an interesting transient in their own right. \citet{Bobrick2022} performed hydrodynamic simulations of mergers of ONe white dwarfs and black holes (and neutron stars), similar to the mergers identified in our \texttt{CMC} models. Focusing on sources in the Galactic field (rather than in clusters), this study found that the nuclear yields of these mergers may contribute significantly to Galactic chemical evolution. They predict transients peaking in the red/infrared that reach peak bolometric magnitudes up to $-16.5$ and evolving on timescales of roughly a month. They argue these events may be similar to faint Type Ia supernovae \citep[e.g.,][]{Jha2006,Phillips2007} and should be readily detectable in large numbers by the Vera Rubin Observatory. Detection of highly off-nuclear events of this type in nearby quiescent galaxies may potentially hint at black hole+white dwarf mergers.

\acknowledgments

We thank Tom Maccarone, Steve Zepf, and Kwangmin Oh for helpful discussions throughout the preparation of this paper. This research was supported in part by grant NSF PHY-2309135 to the Kavli Institute for Theoretical Physics (KITP). This project was completed as a part of the Carnegie Astrophysics Summer Student Internship (CASSI) at The Observatories of the Carnegie Institution for Science and the Summer Undergraduate Research Fellowship (SURF) of the California Institute of Technology under the direct mentorship of Kyle Kremer and the guidance of various scientists at The Observatories, where funding was jointly provided by CASSI and SURF. 
This research was supported in part through the computational resources and staff contributions provided by the Quest high-performance computing facility at Northwestern University.
This work made use of the SPLASH visualization software \citep{2007PASA...24..159P}.

\bibliographystyle{aasjournal}
\bibliography{mybib.bib}

\end{document}